\documentclass[aps,pre,floatfix,twocolumn,showpacs]{revtex4}

\usepackage{graphicx,epsfig}
\usepackage{rotate}

\usepackage{times}
\usepackage{graphics,dcolumn,bm,fleqn,epic,eepic,float}
\usepackage{amssymb,amsmath,multirow,rotate,color}

\newcommand{\be}{\begin{equation}}
\newcommand{\ee}{\end{equation}}

\begin{document}

\title{Cooperation in the Prisoner's Dilemma  game in Random Scale-Free Graphs}

\author{Julia Poncela}

\affiliation{Institute for Biocomputation and Physics of Complex
Systems (BIFI), University of Zaragoza, Zaragoza E-50009, Spain}

\author{Jes\'{u}s G\'{o}mez-Garde\~{n}es}\thanks{To whom correspondence should be addressed. E-mail: gardenes@gmail.com}

\affiliation{Institute for Biocomputation and Physics of Complex
Systems (BIFI), University of Zaragoza, Zaragoza E-50009, Spain}

\affiliation{Departament d'Enginyeria Inform\'atica i Matem\'atiques,
University Rovira i Virgili, Tarragona E-43007, Spain}

\affiliation{Scuola Superiore di Catania,
University of Catania, Catania I-95123, Italy}

\author{Yamir Moreno}

\affiliation{Institute for Biocomputation and Physics of Complex
Systems (BIFI), University of Zaragoza, Zaragoza E-50009, Spain}

\affiliation{Department of Theoretical Physics, University of Zaragoza, Zaragoza E-50009, Spain}

\author{Luis Mario Flor\'{\i}a}

\affiliation{Institute for Biocomputation and Physics of Complex
Systems (BIFI), University of Zaragoza, Zaragoza E-50009, Spain}

\affiliation{Department of Condensed Matter Physics, University of Zaragoza, Zaragoza E-50009, Spain}

\date{\today}

\begin{abstract}

In this paper we study the cooperative behavior of agents playing the Prisoner's Dilemma 
game in random scale-free networks. We show that the survival of
cooperation is enhanced with respect to random homogeneous graphs but, on the other
hand, decreases when compared to that found in Barab\'asi-Albert scale-free
networks. We show that the latter decrease is related with the structure of
cooperation. Additionally, we present a mean field approximation for studying
evolutionary dynamics in networks with no degree-degree correlations and with
arbitrary degree distribution. The mean field approach is similar to the one
used for describing the disease spreading in complex networks, making a
further compartmentalization of the strategists partition into degree-classes. 
We show that this kind of approximation is suitable to describe the behavior of the 
system for a particular set of initial conditions, such as the placement of cooperators 
in the higher-degree classes, while it fails to reproduce the level of cooperation observed 
in the numerical simulations for arbitrary initial configurations.

\end{abstract}


\maketitle

\section{Introduction}

The ubiquity of cooperation at all scales of life's organization from genes,
through multicellular organisms, to animal and human societies, has not been 
immediately accommodated in the theory of evolution [Nowak, 2006]. In fact, 
to understand the observed survival of cooperative behavior when selfish 
actions provide a higher reproductive success (fitness) several mechanisms (non mutually
exclusive) have been proposed: a) Kin selection, based on genetic relatedness,
b) group selection, in which the demes (groups) instead of organisms are the
entities on which natural selection acts upon, and c) reciprocal altruism,
which includes direct reciprocity (through repeated interactions) and diverse
types of indirect reciprocity (through reputation, scoring, punishment,
signaling, etc...).

Perhaps the simplest ({\em i.e.} less demanding) mechanism, termed lattice (or
network) reciprocity, consists of assuming that each agent interacts only with
its neighbors as given by a network of social contacts. Early pioneering
numerical work by Nowak and May [Nowak \& May, 1992] on the evolutionary
dynamics of the Prisoner's Dilemma game (see below) in two-dimensional regular
(square) lattices, showed that the cooperative phenotype was not driven to
extinction. This result stimulated much work paying attention to the evolution
of cooperation in graph-structured populations [Szab\'o \& F\'ath, 2007]. It
has been well recognized that graph topologies play a crucial role in
providing positive feedback evolutionary mechanisms that facilitate the
asymptotic survival of cooperation. Therefore, it is necessary to change the
benchmark of evolutionary game theory from unstructured or ``{\em well-mixed}
populations to the more realistic case of complex networks.

Interactions in animal and human social systems are, in general, modelled on
 sets of individuals playing diverse {\em social dilemmas} games. In these
 games, players can adopt two possible strategies: cooperation (C) and
 defection (D). In the classical setting, a player $i$ plays with all the
 other individuals in the system accumulating, in each of these games, a
 payoff that depends on both its strategy and the one adopted by its
 corresponding opponent. In particular, both C and D receive $R$ (reward)
 under mutual cooperation and $P$ (punishment) under mutual defection, while C
 receives $S$ (suckers) when confronted to D, which in turn receives $T$
 (temptation). The specific values that take the latter payoffs define the
 specific social dilemma we are dealing with. From now on, we will consider
 the case $T>R>P>S$ so that we will focus on the {\em Prisoner's Dilemma} game
 [Hamilton, 1964; Axelrod \& Hamilton, 1981; Hofbauer \& Sigmund, 1998; Hofbauer
 \& Sigmund, 2003; Nowak \& Sigmund, 2005].

Once every agent has played the game with the rest of the population, {\em
i.e.} has completed a game's round-robin, selection takes place and each
player is allowed to change its strategy.  Following the replicator dynamics
[Hofbauer \& Sigmund 1998, Gintis 2000], the probability that an agent adopt a
different strategy in the next generation depends on the difference between
its payoff and the average payoff of the system. Under these conditions
(well-mixed hypothesis or mean field approximation), the fraction of
cooperators will unavoidably decrease in time towards zero. Therefore, it
seems clear that in order to answer the question about how cooperative
behavior can survive in animal and human social systems one has to relax the
hypothesis considered above, such as the well-mixing assumption.

In the last decade, scientist have unveiled the structure of many complex
systems, and have described them in terms of complex networks [Bornholdt \&
Schuster, 2003; Dorogovtsev \& Mendes, 2003; Newman, 2003; Boccaletti, Latora, Moreno, 
Chavez \& Hwang, 2006]. These networks are the backbone of complex systems 
and hence they are the substrates on top of which a number of relevant dynamics 
(such as disease spreading, information transmission or human traits) occur. The 
structure of complex networks is far from being described with a fixed value $k$ 
accounting for the typical number of interactions that an element shares with the 
rest of the system. On the other hand, it has been measured that the distribution of
the number of contacts (or alternatively the degree) of the elements follows a
power-law, $P(k)\sim k^{-\gamma}$, {\em i.e.}  the networks are scale-free
(SF). Moreover, many measures of real complex networks have obtained an
exponent $2<\gamma<3$ for the power-law degree distribution, pointing out that
the variance $\langle k^2\rangle$ of this statistical distribution diverges in
the infinite population limit. Taking into account these results, it becomes
clear that any mathematical model or numerical simulation aimed at describing
the cooperative behavior in real social systems has to incorporate the
scale-free nature of the social contact network.

In this paper we will study the cooperation in random scale-free
networks. First, in section \ref{sec:2}, we will analyze the evolution and the
structure of cooperation in the prisoner's dilemma game through numerical
simulations of the evolutionary dynamics. In particular, we will study the
average level of cooperation sustained by random SF networks and how
cooperators and defectors are arranged across the network. We will compare our
results with those obtained by previous studies on other, scale-free or not,
network topologies. In the second part of the paper, section \ref{sec:3}, we
present a mean field approximation that divides players into classes
accordingly to their number of contacts. The mean field approximation is shown
to be correct for particular sets of initial conditions and hence provides a
useful analytical tool for studying the cooperation in complex networks (in
particular for those graphs with a long-tailed degree distribution). Finally,
in section \ref{sec:4}, we round off the paper by summarizing the main results
of the work.

\section{The structure of cooperation in random scale-free networks}
\label{sec:2}

In the recent years, a number of studies have studied the prisoner's dilemma
on top of complex networks [Egu\'{\i}luz, Zimmermann, Cela-Conde \& San Miguel, 2005; 
Lieberman, Hauert \& Nowak, 2005; Santos \& Pacheco, 2005; Ohtsuki, Hauert, 
Lieberman \& Nowak, 2006; Santos \& Pacheco, 2006; Santos, Pacheco \& Lenaerts, 2006; 
G\'omez-Garde\~nes, Campillo, Flor\'{\i}a \& Moreno, 2007; Poncela, G\'omez-Garde\~nes, 
Flor\'{\i}a \& Moreno, 2007; Szab\'o \& Fath, 2007; G\'omez-Garde\~nes, Poncela, Flor\'{\i}a \& 
Moreno, 2008; Szolnoki, Perc \& Danku, 2008; Vukov, Szab\'o \& Szolnoki, 2008]. 
These works have determined that SF networks enhance cooperation
when compared to homogeneous random networks such as Erd\"os-R\'enyi graphs
(described by a Poisson degree distribution, $P(k)\sim {\mbox e}^{\langle
k\rangle}\langle k\rangle^{k}/k!$). However, these works have mainly focused
on SF networks constructed via the celebrated Bara\'basi-Albert (BA) model
[Barab\'asi \& Albert, 1999]. The BA model considers that the network is grown
from an initial core of $m_0$ nodes, incorporating a new node to the network
every time step. Every new node launches $m$ links to the nodes already
present in the growing network following a preferential attachment rule ({\em
i.e.} the probability of receiving a link from the new node is proportional to
the degree of the nodes).

The networks generated using the BA model have a power-law degree distribution
with $\gamma=3$ but, at the same time, they posses important features that
make them different from random SF networks constructed by means of purely
statistical algorithms such as the Molloy-Reed configurational model [Molloy
\& Reed, 1998]. These differences are manifested in the so-called
age-correlations [Dorogovtsev \& Mendes, 2003; Newman, 2003] that have as a
consequence the interconnection of highly-connected elements or hubs. The
links between hubs have been shown to play a crucial role in the survival of
cooperation [Santos \& Pacheco, 2005] since when they are removed the
cooperation level decreases notably (although it remains larger than those
observed in regular and homogeneous graphs).

In fact, a careful inspection of the structure of cooperation in BA
SF networks [G\'omez-Garde\~nes, Campillo, Flor\'{\i}a \& Moreno,  2007] reveals 
that cooperators are arranged in a very cohesive way. In particular, they are glued 
together into a single cooperator cluster sustained by the highly connected nodes 
that always play as cooperators [G\'omez-Garde\~nes, Poncela, Flor\'{\i}a \& Moreno, 
2008]. Our first goal in the study of random SF networks is to check if the deletion of the
hub-to-hub links affects the structure of cooperation observed in BA networks,
explaining qualitatively the drop in the cooperation level as a break down of
the cohesive arrangement of cooperators.


To study the structure of cooperation in random SF networks we have performed
a rewiring of the SF networks obtained by means of the BA model. Following the
scheme shown in [Maslov \& Sneppen, 2002] the rewiring process destroys any
kind of correlations present in the original network preserving the degree
sequence of the graph, and thus keeping the same degree distribution
($P(k)\sim k^{-3}$) as in the original BA network. The networks generated in
this work have $N=4000$ and $\langle k\rangle=4$.

Once the network is constructed, we perform the numerical simulation of the
evolutionary dynamics as dictated by the prisoner's dilemma payoff matrix with
$P=S=0$, $R=1$, $T=b>1$, so that we deal with only one control parameter:
the temptation to defect $b$. We start from an initial configuration with equal
number of $C$ and $D$ players ($c=d=N/2$) that are randomly distributed across
the network nodes. At each generation of the discrete evolutionary time $t$,
each agent $i$ plays once with every agent in its neighborhood and accumulates
the obtained payoffs $P_i$. Then, all the players update synchronously their
strategies by the following rules. Each individual $i$ chooses at random a
neighbor $j$ and compares its payoff $P_i$ with $P_j$. If $P_i>Pj$, player
$i$ keeps the same strategy for the next generation. On the other hand, if
$P_j>P_i$, the player $i$ adopts the strategy of its neighbor $j$ for the next
game round robin with probability [Santos \& Pacheco, 2005]:
\begin{equation}
\Pi_{i\rightarrow j}=\beta(P_j-P_i)\;.
\label{eq:1}
\end{equation}
Here, $\beta$ is the characteristic inverse time scale: the larger $\beta$ the
faster evolution takes place. Additionally, we consider
$\beta=(\max\{k_i,k_j\}b)^{-1}$ assuring that $\Pi_{i\rightarrow j}\leq1$.

\begin{figure}[t]
\begin{center}
\epsfig{file=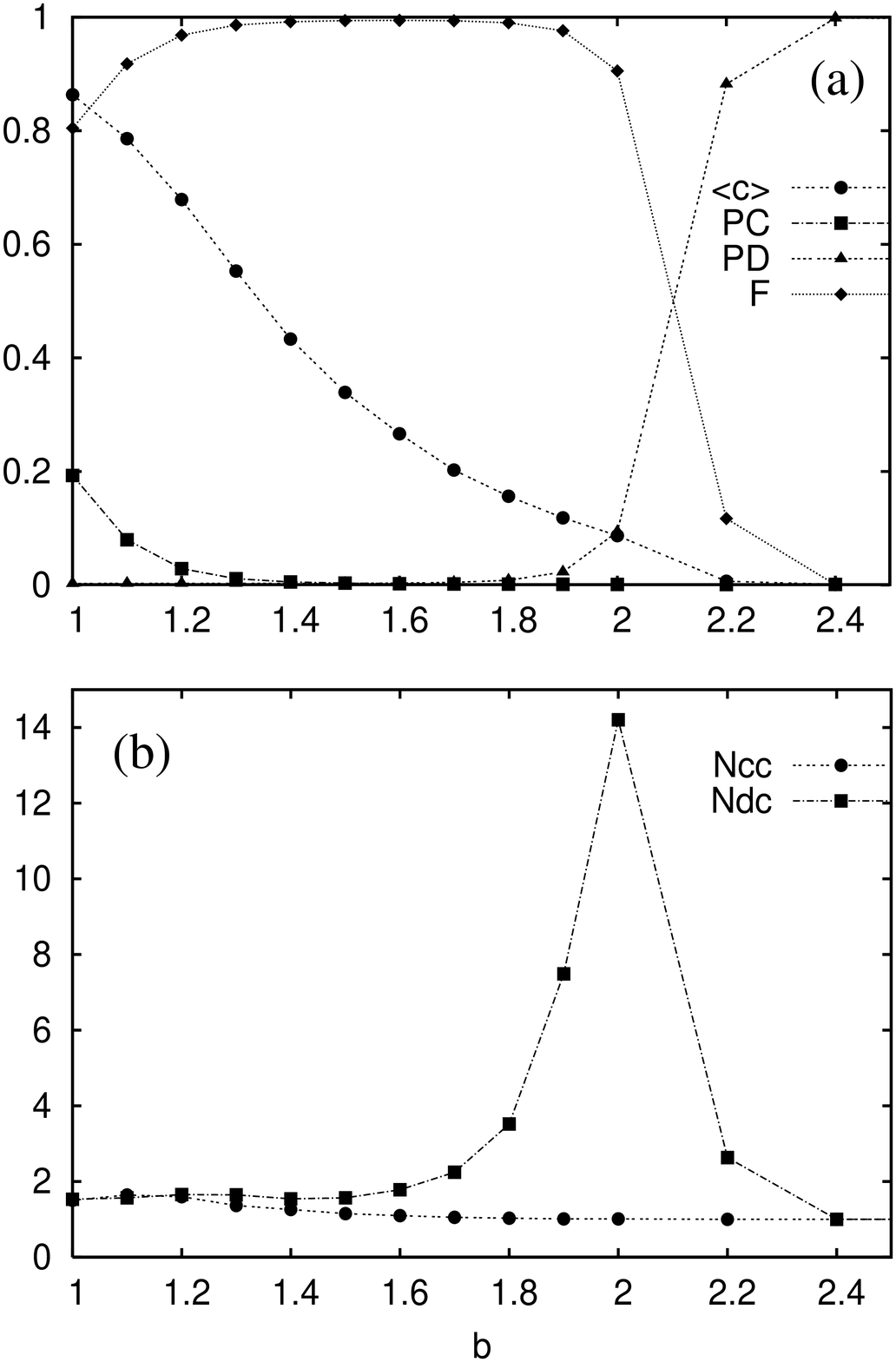,width=3.4in,angle=-0}
\end{center}
\caption{{\bf (a)} Evolution of the average level of cooperation $\langle
  c\rangle$ as a function of the temptation to defect $b$ in random SF
  graphs. The panel also shows the corresponding evolution for the fraction of
  pure cooperators (PC), pure defectors (PD) and fluctuating (F) players. {\bf
  (b)} Evolution of the number of cooperator clusters $N_{cc}$ and defector
  clusters $N_{dc}$ as a function of $b$.}
\label{fig:1}
\end{figure}

Let us now explain the details of the numerical analysis. We let the system
evolve until a stationary regime is reached. The stationary regime is
characterized by a stable average level of cooperation $\langle c\rangle$, that
is the fraction of C players in the network, $\langle c\rangle=c/N$. To
compute $\langle c\rangle$ we let the dynamics evolve over a transient time
$\tau_0=5\cdot 10^3$, and we further evolve the system over time windows of
$\tau=10^3$ generations. In each time window, we compute the average value and
the fluctuations of $c(t)$. When the fluctuations are less than or equal to
$1/\sqrt{N}$, we stop the simulation and consider the average cooperation
obtained in the last time window as the asymptotic average cooperation
$\langle c\rangle$. In order to make an extensive sampling of initial
conditions and network realizations we have performed $10^3$ numerical
simulations for each value of the temptation $b$ studied, and averaged
accordingly the values $\langle c\rangle$ found in the realizations.  In
figure \ref{fig:1}.a we have plotted the evolution of the average level of
cooperation $\langle c\rangle$ as a function of $b$. Our results confirm the
findings of [Santos \& Pacheco, 2005]: the removal of age-correlations makes SF
networks less robust to defection than BA networks. However, the figure shows
that $\langle c\rangle(b)>0$ until larger values of $b$ as compared to the
cooperation levels found for homogeneous ER graphs [G\'omez-Garde\~nes, 
Campillo, Flor\'{\i}a \& Moreno, 2007].

To measure the structure of cooperation we follow the approach introduced in
[G\'omez-Garde\~nes, Campillo, Flor\'{\i}a \& Moreno, 2007]. Once the system 
has reached the stationary regime, we let the system evolve again for 
$\tau_s=10^4$ additional time steps, and measure the relevant magnitudes 
for the dynamical characterization of the stationary state. In particular, 
we label each player in one of the following three categories: 
{\em pure cooperators} (PC), {\em pure defectors} (PD) and 
{\em fluctuating} (F) players. Pure cooperator (pure defector) strategists 
are those players that adopt cooperation (defection) during all
the $\tau_s$ generations. Conversely, fluctuating players are those agents
that play both as defectors and cooperators at the stationary regime, and
hence change their strategies during the time window of length $\tau_s$. In figure
\ref{fig:1}.a we also plot the fraction of PC, PD, and F players as a function of
the temptation $b$. The evolution of the three fractions follows the expected
behavior: PC decrease with $b$, whereas F players first increase and occupy
macroscopically the network. Finally, F agents are progressively replaced by
PD until all the network is fully occupied by pure defectors. Remarkably, the number
of PC is surprisingly lower than both BA and homogeneous ER networks. Instead,
in random SF graphs the average level of cooperation is sustained by those F
players that dominate the population of the network in the range $1.2<b<1.9$.

Once we have identified the PC, PD and F players we define the cooperator and
defector clusters. A cooperator (defector) cluster is a connected subgraph composed of
nodes that are pure cooperators (defectors) and the links between them.  In
figure \ref{fig:1}.b we have plotted the number of cooperators and defectors
clusters as a function of $b$. The first difference with respect to BA networks is
that here we find realizations with more than one cooperator cluster. This
difference explains the drop in the cooperation level observed in [Santos \&
Pacheco, 2005]: the more fragmented the cooperators are arranged the less
sources of benefits they find in their surroundings and the larger is the probability
to be invaded by the instantaneous defectors that are in contact with them.
Regarding the defector clusters we observe the same picture as in BA networks:
PD are arranged in several clusters when they start to invade the network
($b\simeq 2$). The number of defector clusters decrease as they start to grow
in size and finally collapse into a single one only when all the
network is totally invaded by pure defectors.
We have also checked the probability that a node of degree $k$ is a cooperator
in the stationary regime. Our numerical simulations show that, in agreement
with previous numerical observations in BA networks [G\'omez-Garde\~nes, Poncela, 
Flor\'{\i}a \& Moreno, 2008], high degree nodes are more likely to act as cooperators than
intermediate or low degree individuals.

Summing up the previous results, in random SF networks, the fragmentation of the 
cooperator clusters together with the extremely low fraction of pure cooperators 
and the prevalence of fluctuating individuals lead to an organization of cooperation 
that is radically different to that observed in BA SF networks.

\section{The degree-based mean field approximation}
\label{sec:3}

The random SF graphs analyzed above are free of any kind of correlation
between the properties (age, degree, etc...) of two adjacent nodes. Therefore,
it is amenable to study analytically the evolution of the cooperation by
considering a similar approach to that used for diffusion processes in complex
networks [Pastor-Satorras \& Vespignani, 2001a; Pastor-Satorras \& Vespignani,
2001b; Moreno, Pastor-Satorras \& Vespignani, 2002] with arbitrary degree 
distributions and no correlations.  To incorporate the heterogeneity in the number 
of social contacts of individuals we make a further compartmentalization of the
strategists in degree-classes. In this sense, we label $c_k$ and $d_k$ as the
fraction of cooperators and defectors with degree $k$ so that the total number
of cooperators and defectors is, respectively
\begin{eqnarray}
c&=&N\sum_k{P(k)c_k}\;,
\\
d&=&N\sum_k{P(k)d_k}\;.
\end{eqnarray}
Obviously the relation $c_k+d_k=1$ holds and therefore we write the evolution
of the fraction of cooperators with degree $k$ as
\begin{equation}
\dot{c_k}=(1-c_k)\Pi_{k}^{DC}-c_k\Pi_{k}^{CD}\;,
\label{eq:EVOL}
\end{equation}
where $\Pi_{k}^{DC}$ ($\Pi_{k}^{CD}$) is the probability that a cooperator
(defector) of degree $k$ change its strategy to defection
(cooperation). Assuming that the network has no degree-degree correlations,
and following the replicator-like update rule (\ref{eq:1}), we can write the
probabilities $\Pi_{k}^{DC}$ and $\Pi_{k}^{CD}$ as
\begin{eqnarray}
\Pi_{k}^{DC}&=&\sum_{k^{'}}\frac{k^{'}P(k^{'})}{\langle k\rangle}
\beta\;\Theta\left[P_{k^{'}}^C-P_{k}^D\right]c_{k^{'}}\;,
\label{eqDC}
\\
\Pi_{k}^{CD}&=&\sum_{k^{'}}\frac{k^{'}P(k^{'})}{\langle k\rangle}
\beta\;\Theta\left[P_{k^{'}}^D-P_{k}^C\right](1-c_{k^{'}})\;,
\label{eqCD}
\end{eqnarray}
where the function $\Theta[x]$ is defined as $\Theta[x]=x$ if $x>0$ and
$\Theta[x]=0$ otherwise. Besides, $P_{k}^C$ and $P_{k}^D$ are the payoffs
obtained by a cooperator and a defector of degree $k$ respectively, and can be
written as
\begin{eqnarray}
P_{k}^C&=&k\sum_{k^{'}}\frac{k^{'}P(k^{'})}{\langle k\rangle}\;c_{k^{'}}=k{\it
l}_c\;,
\label{eq:Pc}
\\
P_{k}^D&=&b\cdot k {\it l}_c\;,
\label{eq:Pd}
\end{eqnarray}
where ${\it l}_c$ is the probability that a node has a cooperator neighbor. Now
we can insert the above two expressions (\ref{eq:Pc}) and (\ref{eq:Pd}) in
equations (\ref{eqCD}) and (\ref{eqDC}) and finally write the evolution
equation (\ref{eq:EVOL}) as
\begin{eqnarray}
\dot{c_k}&=&
(1-c_k)\sum_{k^{'}>bk}\frac{k^{'}P(k^{'})}{\langle k\rangle}
\beta\;{\it l}_c(k^{'}-bk)c_{k^{'}}
\nonumber
\\
&-&c_{k}\sum_{k^{'}>bk}\frac{k^{'}P(k^{'})}{\langle k\rangle}
\beta\;{\it l}_c(bk^{'}-k)(1-c_{k^{'}})
\nonumber
\\
&-&c_{k}\sum_{k^{'}>k/b}^{bk}\frac{k^{'}P(k^{'})}{\langle k\rangle}
\beta\;{\it l}_c(bk^{'}-k)(1-c_{k^{'}})\;,
\label{eq:9}
\end{eqnarray}
where we have separated the contributions to the transition C$\rightarrow$D
that come from neighbors with $k^{'}>bk$ and $k^{'}<bk$, so that it is clear
that the number of degree classes that participate in the transition
C$\rightarrow$D is larger than those that influence the change
D$\rightarrow$C.


\begin{figure}[!t]
\begin{center}
\epsfig{file=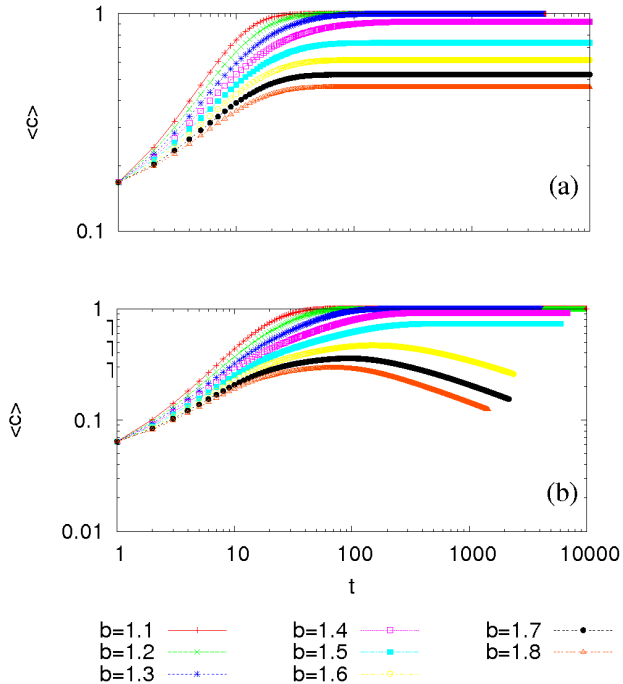,width=3.4in,angle=-0}
\end{center}
\caption{(Color online). The two panels show the time evolution of the
  fraction of cooperators $\langle c\rangle(t)$ obtained solving equation
  (\ref{eq:9}) when targeted cooperation is used as initial conditions and
  being $P(k)$ a power-law with $\gamma=3$. The different curves correspond to
  several values of $b$ (as shown in the bottom of the figure). The targeted
  cooperation used correspond to {\bf (a)} $k^{*}=2$ and {\bf (b)}
  $k^{*}=3$. }
\label{fig:2}
\end{figure}

The main assumption behind the above mean field approach is that the average
level of cooperation inside a degree-class, $c_k$, is a proper magnitude for
describing the state of nodes with this degree. In particular, this assumption
is strictly correct when $c_k$ is either $1$ or $0$. This motivate us to study
the solution of equations (\ref{eq:9}) using a particular set of initial
conditions: the {\em targeted cooperation}.

We define targeted cooperation as a set of initial conditions for of the system
(\ref{eq:9}) for which $c_{k}(0)=1$ if $k> k^{*}$ and $c_{k}(0)=0$ if $k<k^{*}$. We
have carefully explored the solutions of equations (\ref{eq:9}) when $P(k)$ is
a power-law degree distribution. To this end, we have studied power-laws with
several values of $\gamma$ and used different values for the degree threshold
$k^{*}$. The numerical solution of equations (\ref{eq:9}) reveals that the
cooperation level remains for $b>1$, reaching a stationary value that depends
on both the value of $b$ and that of the threshold $k^{*}$. In figure
{\ref{fig:2} we show the time evolution of the average level of cooperation
for several values of $b$ and $k^{*}=2$ and $k^{*}=3$. The degree distribution
in the figure is a power-law with $\gamma=3$. The solutions show that the
larger $k^*$ and/or $b$ are the lower the cooperation level is.

It is interesting to study in detail the effect of the threshold $k^*$ over
the asymptotic level of cooperation. In particular, we can obtain the minimum
amount of degree classes that we have to fill initially with cooperators so
that cooperation is able to survive asymptotically in the network. We have 
explored different sets of initial conditions corresponding to
different values of $k^{*}$. Starting from a low value of $k^{*}$ we have
solved equations (\ref{eq:9}) and computed the final level of cooperation
$\langle c\rangle$. If $\langle c\rangle>0$ we increase the value of $k^*$ and
solve again the system (\ref{eq:9}). This process is iterated until we reach a
value $k^{*}_c$ for which cooperation vanishes. The computed value $k^{*}_c$
thus represents the minimal amount of cooperator degree classes needed at time
$0$ to sustain asymptotically a nonzero level of cooperation. In figure
\ref{fig:3} we have plotted the functions $k^{*}_c(b)$ for three power-law
degree distributions ($\gamma=2$, $3$, $4$). Obviously, we observe that as the
temptation to defect increases it is necessary to fill more degree classes
to assure a nonzero level of cooperation. More interestingly, we show that the
heterogeneity of the network increases the value of $k^{*}_c$. This results is
related to the fact that filling a given amount of degree classes is more
efficient (more nodes are initially set as cooperators) when the network is
more heterogeneous.

\begin{figure}[t!]
\begin{center}
\epsfig{file=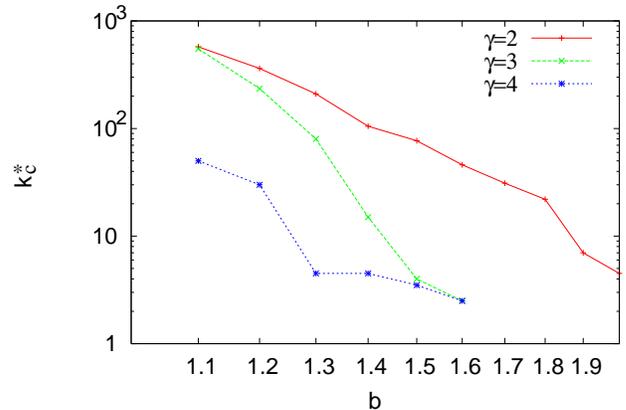,width=3.3in,angle=0}
\end{center}
\caption{Phase diagram $k^{*}_c(b)$. The three curves correspond to different
  power-law distributions (namely, $\gamma=4$, $3$ and $2$). Each curve
  $k^{*}_c(b)$ represent the border between two different asymptotic regimes
  for the evolution of equations (\ref{eq:9}) with targeted cooperation: The
  area below the curves correspond to the points $(b, k^*)$ where targeted
  cooperation yield nonzero asymptotic level of cooperation. Conversely, the
  area above the curves correspond to the targeted initial conditions for
  which the evolution of equations (\ref{eq:9}) yields $\langle
  c\rangle\rightarrow 0$.}
\label{fig:3}
\end{figure}

Computing the phase diagram $k_{c}^{*}(b)$ is a difficult task if we only
rely on the results of the numerical simulations of the evolutionary dynamics
on top of the graphs. Therefore, the mean field approach represents, in this
context, a useful tool for substituting computationally expensive numerical
simulations. However, how accurate are the results of the solutions of
equations (\ref{eq:9}) when compared to numerical simulations?  To check the
reliability of the degree-based mean field approach in the context of targeted
cooperation we have computed the diagram $\langle c\rangle(b)$ for random
SF networks with $\gamma=3$ using two different sets of initial
conditions corresponding to $k^*=3$ and $4$. In figure \ref{fig:4} we show the
results of the numerical simulations compared to the results obtained by
solving equations (\ref{eq:9}). Obviously, the agreement is not complete but
the evolution of the cooperation as a function of $b$ follows the same qualitative behavior
and the cooperation tends to zero ($\langle c\rangle\gtrsim 0$) around the same
values of $b$.
The values of $b$ for which $\langle c\rangle=0$ in each of the curves of the
figures are obviously related to the values $k_c^*$.  Our results show that the 
curves $\langle c\rangle(b)$ obtained from numerical simulations
reach larger values of $b$ with $\langle c\rangle>0$. On the other hand, the
numerical simulations yield very low values of $\langle c\rangle$ for those
values of $b$ for which cooperation asymptotically vanishes solving equations
(\ref{eq:9}). Therefore, the mean field approach seems to be of help to
study the behavior of $k_c^*(b)$ and the asymptotic level of cooperation of
the system when targeted cooperation is initially placed in the system.

\begin{figure}[t!]
\begin{center}
\epsfig{file=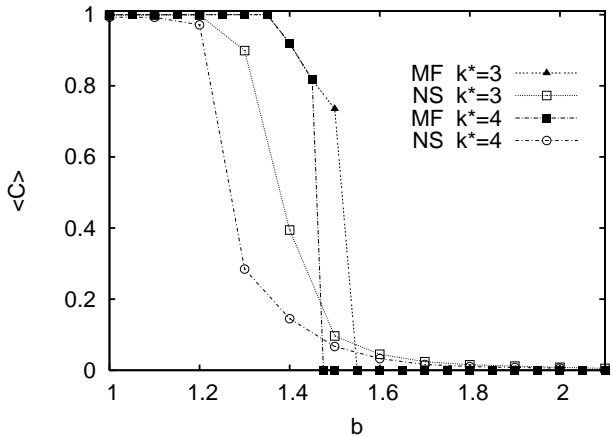,width=3.4in,angle=0}
\end{center}
\caption{Evolution of the asymptotic level of cooperation $\label c\rangle$
  obtained when ({\em i}) solving the mean field (MF) equations (\ref{eq:9})
  and ({\em ii}) computed through numerical simulations (NS) of the
  evolutionary dynamics on top of a random SF network. The degree distribution
  used is a power-law with $\gamma=3$. In both cases we have set targeted
  cooperation as initial conditions for the evolutionary dynamics. We have
  used $k^{*}=3$ and $4$. }
\label{fig:4}
\end{figure}

Regarding general ({\em i.e.} non-targeted cooperation type of)  initial
conditions for the degree-based mean field equations (9), some comments
are in order.  For both, power-law and Poisson  degree distributions  $P(k)$,
random uniformly distributed values for $c_k(0)$, as well as fixed value
$c_k(0) = 0.5$ (mimicking the initial conditions in the numerical simulations
of previous section), led to asymptotic zero level of cooperation as soon as
$b>1$. This suggests that, generically speaking, mean field approaches
(even in generalized forms, as equations (9)) to the evolutionary dynamics
of prisoner's dilemma games on graphs are likely bound to fail to account
for the numerically observed survival of cooperation.  This would be in
agreement with results reported elsewhere [Flor\'{\i}a, Gracia-L\'azaro, 
G\'omez-Garde\~nes \& Moreno, 2008] on a particular type of 
artificial networks that allow a rigorous analysis of the issue. To put 
it in plain terms, the lattice reciprocity mechanisms that enhance the
evolutionary survival of cooperation in network settings seem to be 
out of reach from the (homogeneity) mean field assumptions, in the
sense that they are associated in an essential way to fluctuations of
averaged quantities, like $c_k$ which are the basic descriptors in mean
field approaches.

\section{Conclusions}
\label{sec:4}

Scale-free networks have been recently shown as the graphs that better promote
cooperation. In this article we have shown that the power-law degree
distribution cannot be considered as the only root for the promotion of
cooperation. At variance with the Barab\'asi-Albert networks, the SF graphs
considered in this work are free of any kind of node-node correlation. The
first conclusion of our study is that cooperation decays when no correlations
are present in the network. Moreover, we have shown that the organization of
cooperation is dramatically different from that of the BA network, showing that
cooperators can arrange in more than one cluster increasing the probability of
being invaded by defectors. In other words, the fixation of cooperation is
much lower than in SF networks with correlations, thus completing the picture
provided by other studies where correlations were added into SF networks
[Assenza, G\'omez-Garde\~nes \& Latora, 2008; Push, Weber \& Porto, 2008] 
enhancing the promotion of cooperation of BA networks. Therefore, one one hand, 
our study in random SF networks can be considered as the null model for the 
study of the cooperation in other types of SF graphs. On the other hand, our results
highlight the importance of taking into account other structural properties beyond the 
degree distribution of the network [Da Fontura Costa, Rodrigues, Travieso \& 
Villas Boas, 2007] in order to capture the mechanisms that help to fixate 
cooperation in real complex networks.

The second part of the article  presents a degree-based mean field approach to
study  analitycally  networks  with   arbitrary  degree  distribution  and  no
degree-degree correlations  (such as random SF networks).  The approach relies
in a degree compartmentalization  of cooperators and defectors strategists. We
have  shown  that, contrary to diffusion  dynamics where  a  similar
approach has been applied, the degree-based mean field does not work correctly
when general initial conditions are applied since no  cooperation is observed
when the temptation to defect is larger than the reward to cooperation. On the
other hand, when a particular set of initial conditions is used (consisting in
placing all the  cooperators in the higher degree classes  of the network) the
solution of the mean field yields a non zero level of cooperation for a number
of  targeted  initial configurations.  The  results  obtained  in this  latter
context  qualitatively agree  with  those  obtained  when  extensive  numerical
simulations on top of random SF graphs are performed.

As a conclusion, the results presented in this article complete the studies
about the prisoner's dilemma on top of SF networks showing that node-node
correlations play a key role for sustaining a high level of
cooperation. Besides, the mean field approach, although not working for all
the initial configurations placed in the networks, adds an analytical insight
to the field of evolutionary games on graphs, being especially useful when
targeted cooperation is studied. The mean field approach presented here is
open to the incorporation of other ingredients such as degree-degree
correlations that may help to reproduce the levels of cooperation observed in
the numerical simulations.

\begin{acknowledgments}
  This work has been partially supported by the DGA and the Spanish DGICYT
  Projects FIS2006-12781-C02-01 and FIS2005-00337. J.G.-G. and Y.M. are
  supported by the MICINN through the ``Juan de la Cierva'' and the ``Ram\'on
  y Cajal'' programs.
\end{acknowledgments}


\section{References}

\noindent

Assenza, S., G\'omez-Garde\~nes, J., \& Latora, V. [2008], ``Enhancement of
cooperation in highly clustered scale-free networks'',
\emph{Phys. Rev. E} {\bf 78}, 017101.

Barab\'asi, A.L., \& Albert, R. [1999], ``Emergence of scaling in random
networks'', \emph{Science} {\bf 286}, 509-512.

Boccaletti, S., Latora, V., Moreno, Y., Chavez, M., \& Hwang, D.-U. [2006]
``Complex networks: Structure and dynamics'', \emph{Phys. Rep.} {\bf 424},
175-308.

Bornholdt, S. \& Schuster, H. G. [2003] (Editors) ``Handbook of Graphs and
Networks'' (Wiley-VCH, Germany).

Da Fontura Costa, L., Rodrigues, F. A., Travieso, G., \& Villas Boas, P. R. [2007], 
"Characterization of complex networks: A survey of measurements", 
\emph{Advances in Physics} {\bf 56}, 167-242.

Dorogovtsev, S. N. \& Mendes, J. F. F. [2003] ``Evolution of Networks. From
Biological Nets to the Internet and the WWW'' (Oxford University Press,
Oxford, U.K.).

Egu\'{\i}luz, V.M., Zimmermann, M.G., Cela-Conde, C.J., \& San Miguel,
M. [2005], ``Cooperation and the emergence of role differentiation in the
dynamics of social networks'', \emph{Am. J. Soc.} {\bf 110}, 977-1008.

Flor\'{\i}a, L.M, Gracia-L\'azaro, C., G\'omez-Garde\~nes, J. \& Moreno, Y. [2008], 
"Social Network Reciprocity as a Phase Transition in Evolutionary Cooperation", 
e-print {\it arXiv:0806.4962}.

Gintis, H. [2000], ``Game Theory Evolving'' (Princeton University Press,
Princeton, NJ).

G\'omez-Garde\~nes, J., Campillo, M., Flor\'{\i}a, L.M., \& Moreno, Y. [2007],
``Dynamical organization of cooperation in complex networks'',
\emph{Phys. Rev. Lett.} {\bf 98}, 108103.

G\'omez-Garde\~nes, J., Poncela, J., Flor\'{\i}a, L.M., \& Moreno, Y. [2008],
``Natural selection of cooperation and degree hierarchy in heterogeneous
populations'', \emph{J. Theor. Biol.} {\bf 253}, 296-301.

Hamilton, W.D. [1964], ``The genetical evolution of social behaviour'',
\emph{J. Theor. Biol.} {\bf 7}, 1-16.

Hofbauer, J., \& Sigmund, K. [1998], ``Evolutionary Games and Population
Dynamics'', (Cambrige University Press, Cambridge, UK).

Hofbauer, J., \& Sigmund, K. [2003], ``The evolution of cooperation'',
\emph{Bull. Am. Math. Soc.} {\bf 40}, 479-519.

Lieberman, E., Hauert, C., \& Nowak, M.A. [2005], ``Evolutionary dynamics on graphs'',
\emph{Nature} {\bf 433}, 312-316.

Maslov, S., \& Sneppen, K. [2002], ``Specificity and Stability in Topology of
Protein Networks'' \emph{Science} {\bf 296}, 910-913.

Molloy, M., \& Reed, B. [1998], ``The Size of the Largest Component of a
Random Graph on a Fixed Degree Sequence'', \emph{Combinatorics, Probability
and Computing} {\bf 7}, 295-306.

Moreno, Y., Pastor-Satorras, R., \& Vespignani, A. [2002], ``Epidemic
outbreaks in complex heterogeneous networks'', \emph{Eur. Phys. J. B} {\bf 26}
521-529.

Newman, M.E.J. [2003], ``The structure and function of networks'', \emph{SIAM
Rev.} {\bf 45}, 167-256.

Nowak, M.A. [2006], ``Evolutionary Dynamics. Exploring the Equations of Life'',
(Harvard University Press, Harvard, MA).


Nowak, M.A., \& May, R.M. [1992], ``Evolutionary games and spatial chaos'', \emph{Nature}
{\bf 359}, 826-829.

Nowak, M.A., \& Sigmund, K. [2005], ``The evolution of cooperation''
\emph{Nature} {\bf 437}, 1291-1298.

Ohtsuki, H., Hauert, C., Lieberman, E., \& Nowak, M.A. [2006], ``A simple rule
for the evolution of cooperation on graphs and social networks'',
\emph{Nature} {\bf 441}, 502-505.

Pastor-Satorras, R., \& Vespignani, A. [2001a], ``Epidemic spreading in
scale-free networks'', \emph{Phys. Rev. Lett.} {\bf 86}, 3200-3203.

Pastor-Satorras, R., \& Vespignani, A. [2001b], ``Epidemic dynamics and
endemic states in complex networks'', \emph{Phys. Rev. E} {\bf 63}, 066117.

Poncela, J., G\'omez-Garde\~nes, J., Flor\'{\i}a, L.M., \& Moreno, Y. [2007],
``Robustness of cooperation in the prisoner's dilemma in complex networks'',
\emph{New J. Phys.} {\bf 9}, 184.

Push, A., Weber, S., \& Porto, M. [2008], ``Impact of Topology on the
Dynamical Organization of Cooperation in the Prisoner's Dilemma Game''
\emph{Phys. Rev. E} {\bf 77} 036120.

Santos, F.C., \& Pacheco, J.M. [2005], ``Scale-Free networks provide a unifying
framework for the emergence of cooperation'', \emph{Phys. Rev. Lett.} {\bf
95}, 098104.

Santos, F.C., \& Pacheco, J.M. [2006], ``A new route to the evolution of cooperation'',
\emph{J. Evol. Biol.} {\bf 19}, 726-733.

Santos, F.C., Pacheco, \& J.M., Lenaerts, T. [2006], ``Evolutionary dynamics of social
dilemmas in structured heterogeneous populations'', \emph{Proc. Natl Acad. Sci. USA}
{\bf 103}, 3490-3494.

Szab\'o, G., \& F\'ath, G. [2007], ``Evolutionary games on graphs'',
\emph{Phys. Rep.} {\bf 446}, 97-216.

Szolnoki, A., Perc, M., \& Danku, Z. [2008], ``Towards effective payoffs in the prisoner's
dilemma game on scale-free networks'', \emph{Physica A} {\bf 387}, 2075.

Vukov, J., Szab\'o, G., \& Szolnoki, A. [2008], ``Evolutionary Prisoner's Dilemma game on
the Newman-Watts networks'', \emph{Phys. Rev. E} {\bf 77}, 026109.



\end{document}